\documentclass[12pt]{article}
\usepackage[english]{babel}
\usepackage[utf8]{inputenc}
\pdfoutput=1
\usepackage{xr-hyper}
\usepackage{hyperref}
\usepackage{bm}
\usepackage{amssymb,amsmath,amsfonts,amsthm}
\usepackage[letterpaper, margin=1in]{geometry}
\usepackage[makeroom]{cancel}
\usepackage[margins]{trackchanges}
\usepackage{mathtools}

\usepackage{indentfirst}
\usepackage{graphicx}
\usepackage{diagbox}
\usepackage{hyperref}
\usepackage{caption}
\usepackage{subcaption}
\usepackage{subfloat}
\usepackage{comment}
\usepackage{cleveref}

\usepackage[font=small]{caption}
\usepackage{algorithm}
\usepackage{algpseudocode}
\usepackage{blindtext}
\usepackage{titlesec}
\usepackage{natbib}
\usepackage{yhmath, scalerel,stackengine}
\setcitestyle{authoryear, open={(},close={)}}

\title{\bf Physics-Informed Priors with Application to Boundary Layer Velocity}
  \author{Luca Menicali\\
    Department of Applied and Computational Mathematics and Statistics,\\ University of Notre Dame\\
    and\\
    David H. Richter\\
    Department of Civil and Environmental Engineering and Earth Sciences, \\ University of Notre Dame\\
    and \\
    Stefano Castruccio\\
    Department of Applied and Computational Mathematics and Statistics,\\ University of Notre Dame\\}

\addeditor{LM}
\addeditor{SC}

\begin{document}

\maketitle

\begin{abstract}
One of the most popular recent areas of machine learning predicates the use of neural networks augmented by information about the underlying process in the form of Partial Differential Equations (PDEs). These \textit{physics-informed neural networks} are obtained by penalizing the inference with a PDE, and have been cast as a minimization problem currently lacking a formal approach to quantify the uncertainty. In this work, we propose a novel model-based framework which regards the PDE as a prior information of a deep Bayesian neural network. The prior is calibrated without data to resemble the PDE solution in the prior mean, while our degree in confidence on the PDE with respect to the data is expressed in terms of the prior variance. The information embedded in the PDE is then propagated to the posterior yielding physics-informed forecasts with uncertainty quantification. We apply our approach to a simulated viscous fluid and to experimentally-obtained turbulent boundary layer velocity in a water tunnel using an appropriately simplified Navier-Stokes equation. Our approach requires very few observations to produce physically-consistent forecasts as opposed to non-physical forecasts stemming from non-informed priors, thereby allowing forecasting complex systems where some amount of data as well as some contextual knowledge is available.

\noindent%

\end{abstract}

\clearpage

\section{Introduction \& Background}
\label{sec:intro}

Over the last two decades, data-rich problems have become increasingly more relevant due to the extraordinary increase in the volume, velocity, and variety of data available. The simultaneous growth in computational capabilities has also allowed researchers to implement highly complex models, especially nonparametric constructs such as Neural Networks (NN), able to capture complex patterns in the data. This has facilitated new findings such as predicting long-term trends in the El Ni\~{n}o-Southern Oscillation \citep{enso}, constructing optimal wind farms given scattered observational data \citep{lidar}, and understanding complex flows in computational fluid dynamics \citep{cfd}. 

While NNs have unequivocally presented new opportunities for data-rich problems, in their original formulation they discard any contextual information that may be provided by a theoretical model. This is a major limitation of these constructs, as a significant amount of applied problems consists of some data \textit{as well as} some information in the form of a PDE. In their original formulation, NNs are unsuitable to account for additional context, as parameters are learned by minimizing a loss function which is agnostic of the specific problem \citep{NN_history}. A rapidly-growing branch of machine learning has been focusing on NNs that can account for the physical laws governing a given system of interest. These \textit{physics-informed neural networks} (PINNs) incorporate this knowledge as an additional PDE constraint on the objective function, so that the inference steers the model towards predictions that are physically consistent \citep{raissi}. PINNs have been shown to perform better than NNs in many applications where data is difficult to collect but information on the process in the form of a PDE is available  \citep{blood_flow, lidar, smoke_sparse}. 

While PINNs have demonstrated more flexibility than standard NNs, to our knowledge they have always been regarded as algorithmic minimization problems rather than an outcome of an inferential process from a statistical model. As such, a formal approach to quantify uncertainty in PINNs has not yet been agreed upon \citep{erroraware_pinn}. In the context of NNs, uncertainty quantification can be assessed using dropout, a bootstrapping-inspired method \citep{dropout, dropout2}, or in a Bayesian framework by imposing (often vague) prior distributions, thereby introducing parameter uncertainty \citep{bnn_intro}. In the context of PINNs, Bayesian Neural Networks (BNN) have been used to augment PINNs' predictions in the identification of the system inertia and damping in power systems \citep{power_systems} as well as in the estimation of hypocenter location during seismic activity \citep{seismic}. Even though these studies' results do include measures of uncertainty, they do so in a similar fashion as classical data-driven PINN approaches in forward problems and, most importantly, are cast as PDE solvers \citep{b_pinn}. Previous attempts in fact focus on PINN-driven approximations to PDEs to construct Bayesian priors and eventually augment those deterministic approximations \citep{b_pinn, erroraware_pinn}. To our knowledge, a unified model-based approach aimed at presenting PINNs as outcomes from a statistical model has not been developed yet.

In this work, we propose a novel Bayesian approach that incorporates contextual knowledge in the prior stage of a deep BNN. This approach is fundamentally different from the classical PINN paradigm, which instead adapts the loss function to a set of governing equations \citep{raissi}, and from previous Bayesian-inspired approaches to PINNs, whose main focus remains uncertainty quantification of latent PDE solutions \citep{b_pinn, erroraware_pinn}. At the core of our proposed method is the acknowledgement that an established theory effectively \textit{constitutes} our prior beliefs on a given system's behavior, which is then updated with some data, and is thus an inherently Bayesian approach. This work also addresses the issue of uncertainty quantification in PINNs which is naturally derived from the posterior distribution. In practice, we propose to calibrate the parameters' prior distribution such so the prior mean is close to the PDE solution, while the prior variance reflects our degree of confidence in the PDE with respect to the data. From an epistemological perspective our approach regards PINNs as a Bayesian construct, where the prior calibration occurs \textit{before} any observation is made, so that prior and data are modeled \textit{independently} of each other to obtain physically-consistent posteriors with uncertainty quantification. The large parameter space of BNNs implies a computationally challenging inference, so we approximate it with a variational inference approach \citep{bnn_intro}. Our methodology is very general as it can be adapted to any system for which a NN is to be applied along with some contextual information in the form of a PDE and prescribes a formal approach to incorporate both in the analysis.

We consider an application of a turbulent boundary layer in a water tunnel --- the flow which develops as fluid flows over a wall beneath. The governing set of PDEs, the Navier-Stokes (NS) equations, dictate the conservation of mass and momentum in three spatial dimensions and are generally able to describe accurately the turbulent, time-dependent velocity in boundary layers. For such turbulent flows, however, the goal is rarely to predict the instantaneous velocity field. Instead, the equations are typically averaged so as to predict the \textit{mean} velocity field, which requires certain modeling assumptions. We show how physics-informed priors acknowledge some degree of variation of the mean water velocity, as expected from the experimental configuration. When augmented with a sufficiently flexible prior variance and observations, our approach produces realistic forecasts even with a small number of data. Non-informed priors instead result in non-physical predictions. 

The remainder of the paper is organized as follows. Section \ref{sec:methods} outlines the proposed Bayesian model and the algorithm chosen to estimate its parameters. Section \ref{sec:sim_study} presents a simulation study with a viscous fluid to compare our approach with a standard BNN with vague priors. Section \ref{sec:ns_app} reports the proposed model's results when implemented on the boundary layer velocity application. Section \ref{sec:conclusion} summarizes the paper's findings and discusses future work. 

\section{Methods}
\label{sec:methods}

This section formally introduces the general Bayesian framework and proceeds as follows. We first review deep BNNs in Section \ref{sec:bnns}. Then, we introduce our novel BNN prior physical calibration in Section \ref{sec:pinn_background}. Due to the high number of parameters in the BNN, deriving the posteriors is computationally challenging, so Section \ref{sec:post_inf} presents a variational inference approach to approximate the posteriors.

\subsection{Bayesian Neural Network}
\label{sec:bnns}
We consider a spatio-temporal process $u(\bm{s},t)$, where    $(\bm{s},t) \in \Omega \times [0,T] \subset \mathbb{R}^{d} \times [0, \infty)$,
and $d$ is the number of spatial dimensions. We model $u$ with a $L$-layer deep BNN parameterized by $\bm{\theta}=(\bm{\theta}_{1}, \ldots,\bm{\theta}_{L}) \in \Theta$, where $\bm{\theta}_{l}=\{\bm{W}_{l},\bm{b}_{l}\}$ are the parameters in the $l$-th layer, comprised of its weights $\bm{W}_{l}$ and bias terms $\bm{b}_{l}$, and $\Theta$ is the parameter space. Formally, we have
\begin{subequations} \label{eq:bnn}
\begin{align}
    \bm{H}_{0} &= (\bm{s},t),  \label{eq:bnn1} \\
    \bm{H}_{l} &= g(\bm{W}_{l}\bm{H}_{l-1} + \bm{b}_{l}),~ l \in \{1,\ldots,L\},  \label{eq:bnn2} \\
     \bm{H}_{L} &= \left(\mu_{u}(\bm{s},t), \sigma^{2}_{u}(\bm{s},t) \right)\in \mathbb{R} \times [0, \infty), \label{eq:bnn3} \\ 
    u(\bm{s},t) &\sim \mathcal{N}\left( \mu_{u}(\bm{s},t), \sigma_{u}^{2}(\bm{s},t) \right), \label{eq:bnn5}
\end{align} 
\end{subequations}
where the $L$-th layer is two-dimensional, every hidden layer $l \in \{1,\ldots,L-1\}$ has the same number of nodes $K$ (although this assumption can be relaxed), and $u(\bm{s},t)$ is distributed as a normal distribution with mean $\mu_{u}(\bm{s},t)$ and variance $\sigma_{u}^{2}(\bm{s},t)$. Each realization of $u$ is therefore assumed to be conditionally independent given the mean and variance. The non-linear activation function $g$ is assumed to be the same for all $l$, and common choices for $g$ are the rectified linear unit, sigmoid, or hyperbolic tangent function \citep{goodfellow2016}.

The parameters are assumed to have prior distribution 
\begin{align}
    p(\bm{\theta}_{l}) = \mathcal{N}(\bm{\mu}_{l}, \bm{\Sigma}_{l}),~ l \in \{1,\ldots,L\}. \label{eq:priors_gen}
\end{align} 
Then, if we denote with $\bm{u}=\{u(\bm{s}_{1},t_{1}),\ldots, u(\bm{s}_{n},t_{n})\}$ the observations collected at $n$ spatio-temporal locations and $p(\bm{\theta}) =  \prod_{l=1}^{L} p(\bm{\theta}_{l})$ the prior of $\bm{\theta}$ (i.e., $\bm{\theta}_{l}$ are a priori independent), then we can write the posterior as $p(\bm{\theta} \mid \bm{u}) \propto p(\bm{u} \mid \bm{\theta})p(\bm{\theta})$,
where $p(\bm{u} \mid \bm{\theta})$ is the Gaussian likelihood implied by \eqref{eq:bnn}. While $p(\bm{\theta} \mid \bm{u})$ is formally straightforward, the dimensionality of $\bm{\theta}$ comprising, among others, of all entries in the weight matrices $\bm{W}_{l}$, is such that in practice this expression is computationally intractable. In fact, except for some degenerate cases, e.g., a single-layer BNN with an identity activation function that degenerates into a Bayesian linear regression model, conjugate priors in deep BNNs do not generally exist and the large parameter space makes posterior sampling numerically infeasible. For this reason, variational inference techniques are used to approximate the true posterior and generate forecasts \citep{bnn_intro}, as we show in Section \ref{sec:post_inf}. 

\subsection{Physics-Informed Priors}
\label{sec:pinn_background} 

The focus of this work lies in the choices of prior mean $\bm{\mu}=(\bm{\mu}_{1},\ldots,\bm{\mu}_{L})$ and variance $\bm{\Sigma}=(\bm{\Sigma}_{1},\ldots,\bm{\Sigma}_{L})$ in \eqref{eq:priors_gen} so that the priors are informed by the physics of the problem. To this end, we assume the existence of a theoretical model in the form of a spatio-temporal PDE for $u$: 
\begin{subequations} \label{eq:pde}
\begin{alignat}{2}
    \frac{\partial u(\bm{s},t)}{\partial t}+\mathcal{N}[u(\bm{s},t)]&=0,~ &&\bm{s} \in \Omega,~ t \in [0,T], \label{eq:pde_gen} \\
    u(\bm{s},0) &= h_{\text{IC}}(\bm{s}),~&&\bm{s} \in \Omega, \label{eq:pde_ic} \\
    u(\bm{s},t) &= h_{\text{BC}}(\bm{s},t),~&&\bm{s} \in \partial \Omega,~t \in [0,T], \label{eq:pde_bc}
\end{alignat}
\end{subequations}
where $\mathcal{N}[\cdot]$ is a (possibly non-linear) differential spatial operator that may contain derivatives with respect to $\bm{s}$, \eqref{eq:pde_ic} is the initial condition, \eqref{eq:pde_bc} is the boundary condition, and $\partial \Omega$ is the boundary of $\Omega$. Our objective is to calibrate the parameters $\bm{\mu}$ so that a prior for $u$ captures as much information carried by the PDE \eqref{eq:pde} as possible.

A deep NN such as \eqref{eq:bnn} can be augmented to yield prior means that are compliant with physical laws \citep{raissi}. Let $\hat{u}_{\bm{\mu}}(\bm{s},t)$ be the estimate of $u(\bm{s},t)$ as induced by $\bm{\mu}$, the prior mean of  ${\bm{\theta}}$ at spatio-temporal location $(\bm{s},t)$. Well-calibrated prior means are such that $\hat{u}_{\bm{\mu}}(\bm{s},t)$ conforms with the physical model \eqref{eq:pde}. Given any input $(\bm{s},t)$, we can find values of $\bm{\mu}$ that minimize
\begin{align}
    \hat{f}_{\bm{\mu}}(\bm{s},t) := \frac{\partial \hat{u}_{\bm{\mu}}(\bm{s},t)}{\partial t}+\mathcal{N}[\hat{u}_{\bm{\mu}}(\bm{s},t)], \label{eq:f_hat}
\end{align}
while also being compliant with the initial conditions in \eqref{eq:pde_ic} and boundary conditions in \eqref{eq:pde_bc}. We discretize the domain over which \eqref{eq:pde} exists with $N_{\text{PDE}}$, $N_{\text{IC}}$, and $N_{\text{BC}}$ points, representing the number of grid points in $\Omega \times [0,T]$, $\Omega \times \{0\}$, and in $\partial \Omega \times [0,T]$, identifying each of the PDE conditions \eqref{eq:pde_gen}-\eqref{eq:pde_bc}, respectively. Then, we can obtain $\bm{\mu}$ as
\begin{align} 
    \bm{\mu} := \hat{\bm{\mu}}_{\text{physics}} &= \mathop{\arg \min} \limits_{\bm{\mu}} \text{MSE}(\bm{\mu}) \nonumber \\ 
    &= \mathop{\arg \min} \limits_{\bm{\mu}} \left\{ \text{MSE}_{\text{PDE}}(\bm{\mu}) + \text{MSE}_{\text{IC}}(\bm{\mu}) + \text{MSE}_{\text{BC}}(\bm{\mu}) \right\}  \nonumber \\
    &= \mathop{\arg \min} \limits_{\bm{\mu}} \Bigg\{ \int_{\Omega \times [0,T]} \left|\hat{f}_{\bm{\mu}}\left(\bm{s},t\right) \right|^{2} \text{d}\bm{s} ~  \text{d} t +  \int_{\Omega} \left|h_{\text{IC}}\left(\bm{s}\right)-\hat{u}_{\bm{\mu}}\left(\bm{s},0\right) \right|^{2}~\text{d} \bm{s} \nonumber \\  
    &\qquad + \int_{\partial \Omega \times [0,T]} \left|h_{\text{BC}}\left(\bm{s},t\right)-\hat{u}_{\bm{\mu}}\left(\bm{s},t\right) \right|^{2}~\text{d} \bm{s}~\text{d} t  \Bigg\} \nonumber \\
    &\approx \mathop{\arg \min} \limits_{\bm{\mu}} \Bigg\{ \frac{1}{N_{\text{PDE}}} \sum_{i=1}^{N_{\text{PDE}}} \left|\hat{f}_{\bm{\mu}}\left(\bm{s}_{i},t_{i}\right) \right|^{2} + \frac{1}{N_{\text{IC}}} \sum_{i=1}^{N_{\text{IC}}} \left|h_{\text{IC}}\left(\bm{s}_{i}\right)-\hat{u}_{\bm{\mu}}\left(\bm{s}_{i},0\right) \right|^{2}  \nonumber \\ 
    &\qquad + \frac{1}{N_{\text{BC}}} \sum_{i=1}^{N_{\text{BC}}} \left|h_{\text{BC}}\left(\bm{s}_{i},t_{i}\right)-\hat{u}_{\bm{\mu}}\left(\bm{s}_{i},t_{i}\right) \right|^{2}  \Bigg\}. \label{eq:mse}
\end{align}

In summary, we assign values to the prior means $\bm{\mu}$ as determined by the minimizers of $\eqref{eq:mse}$, so that the prior is compliant with the physical model \eqref{eq:pde}. It is important to note that we are \textit{not} interested in solving the PDE \eqref{eq:pde}. Rather, this formulation yields prior means $\hat{\bm{\mu}}_{\text{physics}}$ that carry physical meaning, specifically one that expresses actual prior beliefs, i.e., our best guess for the observed behavior of $u$ resembling the physical model \eqref{eq:pde}.

The prior variance $\bm{\Sigma}_l$ in \eqref{eq:priors_gen}, $l=1,\ldots, L$, is a \textit{de-facto} measure of confidence in the prior means $\hat{\bm{\mu}}_{\text{physics}} 
=(\hat{\bm{\mu}}_{1,\text{physics}},\ldots, \hat{\bm{\mu}}_{L,\text{physics}})$. Here we assume $\bm{\Sigma}_{l,ij} = 0$ for $i \neq j$, i.e., a priori independence for the elements of $\bm{\theta}_{l}$ for all $l$, and denote the diagonal of $\bm{\Sigma}_{l}$ by $\bm{\sigma}^{2}_{l,\text{physics}}$, where $\bm{\sigma}^{2}_{\text{physics}}=(\bm{\sigma}^{2}_{1,\text{physics}},\ldots,\bm{\sigma}^{2}_{L,\text{physics}})$. For a given $C>0$, expressing our relative degree of confidence in the theoretical model against the data, and $N_{l}$, the number of elements in $\hat{\bm{\mu}}_{l,\text{physics}}$, we let 
\begin{align*}
    \bm{\sigma}^{2}_{l,\text{physics}} := \left(\frac{C}{N_{l}} \hat{\bm{\mu}}_{l,\text{physics}}\right)^{2},~l \in \{1,\ldots,L\},
\end{align*}
so that the variance of each $\bm{\theta}_{l}$ is relative to its mean $\hat{\bm{\mu}}_{l,\text{physics}}$ and, from \eqref{eq:priors_gen} we have
\begin{align}
    p\left( \bm{\theta}_{l} \right) = \mathcal{N}\left(\bm{\mu}_{l}=\hat{\bm{\mu}}_{l,\text{physics}},~\bm{\Sigma}_{l}=\text{diag}\left(\bm{\sigma}_{l,\text{physics}}^{2}\right) \right),~l \in \{1,\ldots,L\}. \label{eq:prior}
\end{align}
If we let $\hat{u}_{\hat{\bm{\mu}}_{\text{physics}}}(\bm{s},t)$ be the estimate of $u(\bm{s},t)$ as induced by $\hat{\bm{\mu}}_{\text{physics}}$, previous work \citep{erroraware_pinn} suggests a possible choice for $C$ determined by a function of the mean squared error between $\hat{u}_{\hat{\bm{\mu}}_{\text{physics}}}(\bm{s},t)$ and $u(\bm{s},t)$. However, such a metric would employ data and the priors would no longer reflect \textit{prior} beliefs. In general, values of $C$ close to zero imply a high level of confidence in $\hat{\bm{\mu}}_{\text{physics}}$ and, by extension, a posterior that will skew in the direction the physical model (\ref{eq:pde}). On the other hand, large values of $C$ grant the likelihood more flexibility to drive posterior inference, rendering the model similar to one with vague priors. In both cases, the uncertainty encoded in the physics-informed priors propagates through the BNN and the influence that the physics places on the posterior, and ultimately the forecasts, is rooted in an established theoretical model. 

\subsection{Posterior Inference}
\label{sec:post_inf}

In this work, we approximate the true posterior $p(\bm{\theta}\mid \bm{u})$ with variational inference methods, which we summarize here \citep{blundell_bnn, vi_blei, bnn_intro4}. Further technical details on the mathematical formulation are available in the supplementary materials. We approximate the posterior using a distribution $q(\bm{\theta};\bm{\eta}) \in \mathcal{Q}$, where $\mathcal{Q}$ is a family of Gaussian distributions with dimensionality equal to that of $p(\bm{\theta}\mid \bm{u})$. The variational parameters are $\bm{\eta}=(\bm{\mu}_{*}, \bm{\rho}_{*})$, where $\bm{\rho}_{*} = \log\left[ \exp(\bm{\sigma}_{*}) - 1 \right]$ and $q(\bm{\theta};\bm{\eta})$ has mean $\bm{\mu}_{*}$ and diagonal covariance $\text{diag}(\bm{\sigma}_{*}^{2})$.
We want to find the density $\hat{q}(\bm{\theta} ; \hat{\bm{\eta}}) \in \mathcal{Q}$ that is a suitable approximation for $p(\bm{\theta} \mid \bm{u})$, which we accomplish by minimizing the Kullback-Leibler (KL) divergence between the variational and true posterior \citep{kl_1951}. The variational approximation $\hat{q}(\bm{\theta} ; \hat{\bm{\eta}})$ also indirectly depends on the observations $\bm{u}$ through the KL divergence and the shape of $q(\bm{\theta};\bm{\eta})$ is determined by $\bm{\eta}$ \citep{vi_blei}. The optimal choice of $q(\bm{\theta} ; \bm{\eta})$, defined as 
\begin{align}
    \hat{q}(\bm{\theta};\hat{\bm{\eta}}) &= \mathop{\arg \min}\limits_{q \in \mathcal{Q}} \text{KL}(q(\bm{\theta};\bm{\eta}) \mid \mid p(\bm{\theta}\mid \bm{u})) \nonumber \\ 
    &= \mathop{\arg \min}\limits_{q \in \mathcal{Q}}  \int_{\Theta} q(\bm{\theta}; \bm{\eta}) \log \left( \frac{q(\bm{\theta}; \bm{\eta})}{p(\bm{\theta} \mid \bm{u})}\right)~\text{d}\bm{\theta} \nonumber \\
    &= \mathop{\arg \min}\limits_{q \in \mathcal{Q}}  \int_{\Theta} \left[ \underbrace{\log q(\bm{\theta}; \bm{\eta})}_{\substack{\text{Variational} \\ \text{distribution}}} - \underbrace{\log p(\bm{\theta})}_{\substack{\text{Physics} \\ \text{prior}}} - \underbrace{\log p(\bm{u} \mid \bm{\theta})}_{\text{Likelihood}} \right] q(\bm{\theta}; \bm{\eta})~\text{d}\bm{\theta}, \label{eq:kl_true}
\end{align}
where the last equation holds because of the Bayes theorem. The minimization can be achieved to computing gradients of the cost function
\begin{align}
    \mathcal{F}(\bm{\eta}, \bm{u}) &:= \int_{\Theta} \left[ \log q(\bm{\theta}; \bm{\eta}) - \log p(\bm{\theta}) - \log p(\bm{u} \mid \bm{\theta})\right] q(\bm{\theta}; \bm{\eta})~\text{d}\bm{\theta}  \nonumber \\
    &= \text{KL}(q(\bm{\theta};\bm{\eta}) \mid \mid p(\bm{\theta})) - \int_{\Theta} \log p(\bm{u} \mid \bm{\theta}) q(\bm{\theta}; \bm{\eta}) ~\text{d}\bm{\theta}, \label{eq:kl_cost}
\end{align}
with respect to $\bm{\eta}$. While the first integral in \eqref{eq:kl_cost} can be computed analytically, as  $q(\bm{\theta}; \bm{\eta})$ and $p(\bm{\theta})$ are both Gaussian, the second integral is such that minimizing \eqref{eq:kl_cost} na\"ively is computationally demanding \citep{blundell_bnn, vi_trick}. Indeed, standard Monte Carlo approximation, i.e., drawing samples $\bm{\theta}^{(j)}$ from $q(\bm{\theta} ; \bm{\eta})$ directly, $j=1,\ldots,E$, leads to a gradient estimator of the likelihood term in \eqref{eq:kl_cost},
\begin{align*}
    \frac{1}{E} \sum_{j=1}^{E} \log p\left(\bm{u} \mid \bm{\theta}\right) \nabla_{q(\bm{\theta}^{(j)})} \log q\left(\bm{\theta}^{(j)}; \bm{\eta}\right) \approx \nabla_{\bm{\eta}} \left[ \int_{\Theta} \log p(\bm{u} \mid \bm{\theta}) q(\bm{\theta}; \bm{\eta}) ~\text{d}\bm{\theta} \right],
\end{align*}
with very high variance when $E$ is low and significant computational burden when $E$ is high \citep{vi_grad}. Such issues arise since sampling from $q(\bm{\theta} ; \bm{\eta})$ directly would yield samples that depend on $\bm{\eta}$, with respect to which we are trying to optimize. Here we estimate $\hat{q}(\bm{\theta} ; \hat{\bm{\eta}})$ using the Bayes By Backprop (BBB) optimization algorithm, an adaptation of stochastic gradient descent applied to variational inference \citep{blundell_bnn}, which addresses this issue. We describe the approximate minimization of $\mathcal{F}(\bm{\eta}, \bm{u})$ in general, i.e., assuming that the first term in \eqref{eq:kl_cost} may not be computed analytically, even though in our work we use its exact KL divergence.

BBB predicates estimation of $\mathcal{F}(\bm{\eta}, \bm{u})$ (and ensuing gradients) using an auxiliary variable $\bm{\epsilon} \sim q(\bm{\epsilon})=\mathcal{N}(\bm{0},\bm{I}) \in \mathcal{Q}$. Given any (fixed) value of the variational parameters $\bm{\eta}^{(i)}$, we sample $E$ realizations from $q(\bm{\epsilon})$, $\{\bm{\epsilon}^{(1)},\ldots, \bm{\epsilon}^{(E)}\}$, and obtain samples from the variational distribution as 
\begin{align*}
    \bm{\theta}^{(i,j)} := \bm{\mu}^{(i)}_{*} + \log \left(1+\exp\left(\bm{\rho}^{(i)}_{*}\right)\right) \circ \bm{\epsilon}^{(j)},~j=1,\ldots,E,
\end{align*}
where ``$\circ$" indicates point-wise multiplication. We then write the approximation
\begin{align}
    \widehat{\mathcal{F} \left (\bm{\eta}^{(i)}, \bm{u} \right)} &= \left\{ \frac{1}{E} \sum_{j=1}^{E} \left[ \log q\left(\bm{\theta}^{(i,j)}; \bm{\eta}^{(i)} \right) - \log p\left(\bm{\theta}^{(i,j)} \right) - \log p\left(\bm{u} \mid \bm{\theta}^{(i,j)} \right) \right] \right\} \nonumber \\ 
    &= \left\{ \frac{1}{E} \sum_{j=1}^{E} \phi \left(\bm{\theta}^{(i,j)}, \bm{\eta}^{(i)},\bm{u} \right) \right\}. \label{eq:kl_estimator}
\end{align}
Since the function $\phi(\cdot)$ is differentiable and $\bm{\theta}^{(i,j)}$ is not drawn directly from the variational distribution $q(\bm{\theta};\bm{\eta})$, one can show (proof available in the supplementary material) that $\nabla_{\bm{\eta}} \mathbb{E}_{q(\bm{\theta};\bm{\eta})} \left[\phi(\bm{\theta}, \bm{\eta},\bm{u})\right] = \mathbb{E}_{q(\bm{\epsilon})} \left[ \nabla_{\bm{\eta}} \phi(\bm{\theta}, \bm{\eta},\bm{u})\right]$, so we can use the gradient estimator
\begin{align*}
    \widehat{\nabla_{\bm{\eta}} \mathcal{F}\left(\bm{\eta}^{(i)}, \bm{u}\right)} = \left( \frac{1}{E} \sum_{j=1}^{E} \nabla_{\bm{\eta}} \phi\left(\bm{\theta}^{(i,j)},\bm{\eta}^{(i)},\bm{u}\right) \right),
\end{align*}
which is unbiased for $\nabla_{\bm{\eta}} \mathcal{F}\left(\bm{\eta}^{(i)}, \bm{u}\right)$ \citep{blundell_bnn}, to update the variational parameters $\bm{\eta}$ until some convergence criterion is met. The minimization of the original KL divergence in \eqref{eq:kl_true} is now reduced to computing gradients of a differentiable function with respect to $\bm{\eta}$, as in the case of stochastic gradient descent for standard neural networks.

We also set $E=1$ and divide the $n$ observations in $\bm{u}$ into $B$ mini-batches, where each mini-batch $\bm{u}_{b}$ contains $M$ observations (i.e., assume for simplicity $n=MB$), so that $\bm{u}_{b}=\{u_{b,1},\ldots,u_{b,M}\}$ and $\bm{u}=\{\bm{u}_{1},\ldots,\bm{u}_{B}\}$, which is a common strategy to achieve faster convergence while lessening the computational burden \citep{vi_trick}. The mini-batch optimization implemented in this work computes \eqref{eq:kl_estimator} for each $\bm{u}_{b}$ as
\begin{align*}
    \hat{\mathcal{F}}\left(\bm{\eta}^{(i)}, \bm{u}_{b}\right) &= \frac{M}{n} \text{KL}\left(q\left(\bm{\theta};\bm{\eta}^{(i)}\right) \mid \mid p(\bm{\theta})\right)  -  \log p\left(\bm{u}_{b} \mid \bm{\theta}^{(j)}\right),
\end{align*}
so that the variational parameters are updated $BN_{\text{BBB}}$ times during the optimization, where $N_{\text{BBB}}$ denotes the total number of iterations. The entire optimization procedure is summarized in Algorithm \ref{alg:bbb} and further details on the KL divergence are available in the supplementary material.

\begin{algorithm}
  \caption{Bayes By Backprop} \label{alg:bbb}
  \begin{algorithmic}[1]
    \Procedure{BBB}{$B, M, N_{\text{BBB}}, n$}
    \State $i=0$
    \State $\bm{\eta} = (\bm{\mu}_{*}, \bm{\rho_{*}})$ \Comment{Initialize $\bm{\eta}$}
      \For{$i < N_{\text{BBB}}$}
        \State $b=0$
        \For{$b < B$}
        \State Draw $\bm{\epsilon} \sim \mathcal{N}(\bm{0},\bm{I})$
        \State $\bm{\theta} :=  \bm{\mu}_{*} + \log[1 + \exp(\bm{\rho_{*}})] \circ \bm{\epsilon}$  \Comment{Reparameterization}
        \State $\phi(\bm{\theta}, \bm{\eta}) =  \frac{M}{n} \text{KL}\left(q\left(\bm{\theta};\bm{\eta}\right) \mid \mid p(\bm{\theta})\right)  - \frac{1}{M} \sum_{j=1}^{M}  \log p\left(u_{j} \mid \bm{\theta} \right) $ 
        \State $ \nabla_{\bm{\mu}_{*}} \phi =  \frac{\partial \phi(\bm{\theta}, \bm{\eta})}{\partial \bm{\theta}} + \frac{\partial \phi(\bm{\theta}, \bm{\eta})}{\partial \bm{\mu}_{*}}$ \Comment{Proposition S1}
        \State $ \nabla_{\bm{\rho_{*}}} \phi =  \frac{\partial \phi(\bm{\theta}, \bm{\eta})}{\partial \bm{\theta}} \frac{\bm{\epsilon}}{1+\exp(-\bm{\rho_{*}})} + \frac{\partial \phi(\bm{\theta}, \bm{\eta})}{\partial \bm{\rho_{*}}}$ \Comment{Proposition S1}
        \State $\bm{\gamma} = (\nabla_{\bm{\mu}_{*}},  \nabla_{\bm{\rho_{*}}})$ \Comment{Unbiased MC gradients}
        \State $\bm{\eta} = \bm{\eta} - \alpha \bm{\gamma}$ \Comment{$\alpha$ is the learning rate}
        \State $b = b+1$
        \EndFor
        \State $i = i+1$
      \EndFor
      \State \textbf{return} $\hat{\bm{\mu}}_{*}$, $\hat{\bm{\rho}}_{*}$ 
    \EndProcedure
  \end{algorithmic}
\end{algorithm}

The optimization yields the Gaussian distribution $\hat{q}(\bm{\theta} ; \hat{\bm{\eta}})$ that balances the physics-informed priors and the likelihood, resulting in an approximation of the posterior that is informed by the theoretical and observed behavior of $u(\bm{s},t)$. From the estimated $\hat{\bm{\eta}} = (\hat{\bm{\mu}}_{*}, \hat{\bm{\rho}}_{*})$, we retrieve $\hat{\bm{\sigma}}_{*}^{2} = \log(\exp(\hat{\bm{\rho}}_{*}) + 1)^{2}$ and have
\begin{gather*}
    p(\bm{\theta} \mid \bm{u}) \approx \hat{q}(\bm{\theta} ; \hat{\bm{\eta}}) = \mathcal{N}\left(\bm{\hat{\mu}}_{*}, \text{diag}(\hat{\bm{\sigma}}^{2}_{*})\right),
\end{gather*}
so that we can compute and sample from the posterior predictive distribution as
\begin{align}
    p(\mu_{u}(\bm{s},t) \mid \bm{u}) &= \int_{\Theta} p(\mu_{u}(\bm{s},t) \mid \bm{\theta}) p(\bm{\theta} \mid \bm{u}) ~\text{d} \bm{\theta}, \nonumber \\
    &\approx \int_{\Theta} p(\mu_{u}(\bm{s},t) \mid \bm{\theta}) \hat{q}(\bm{\theta} ; \hat{\bm{\eta}}) ~\text{d} \bm{\theta}, \nonumber \\
    &\approx \frac{1}{M} \sum_{i=1}^{M} p(\mu_{u}(\bm{s},t) \mid \bm{\theta}^{(i)}),~\bm{\theta}^{(i)} \sim \hat{q}(\bm{\theta} ; \hat{\bm{\eta}}), \label{eq:post_mean}
\end{align}
and likewise from $p(\sigma^{2}_{u}(\bm{s},t) \mid \bm{u})$, yielding physics-informed forecasts $\mathbb{E}(\mu_{u}(\bm{s},t) \mid \bm{u})$ and physics-informed variance $\mathbb{E}(\sigma_{u}^{2}(\bm{s},t) \mid \bm{u})$ \citep{blundell_bnn}. The prior means $\hat{\bm{\mu}}_{\text{physics}}$ and prior variance $\bm{\sigma}^{2}_{\text{physics}}$ thus influence the posterior mean \textit{and} posterior variance, yielding physics-informed forecasts \textit{and} physics-informed variance.  

\section{Simulation study}
\label{sec:sim_study}

We conduct a simulation study in a system which is fully explained by a physical model. In Section \ref{sec:sim_details} we report the simulation details. In Section \ref{sec:sim_results} we assess the performance of our general Bayesian framework with physics-informed priors against the non-informed alternative. In Section \ref{sec:sim_sensitivity} we conduct sensitivity analyses by considering different scenarios of data availability as well as specifications of the physical model.

\subsection{Simulation details}
\label{sec:sim_details}

\subsubsection*{Governing PDE}
\label{sec:sim_details_model}

We simulate data from a one-dimensional Burgers' equation, a fundamental non-linear PDE in fluid dynamics obtained from NS, describing a one-dimensional viscous fluid's behavior as it reverts back to its resting state after an initial shock \citep{burgers1948}. Formally, the spatio-temporal process $u(\bm{s},t)$ is the velocity of a fluid as a function of space and time, where $d=1$ so that $\bm{s}=x$,
\begin{align}
    (x,t) \in \Omega \times [0,T] = [-1,1] \times [0,2], \label{eq:burgers_domain}
\end{align}
and the governing model is
\begin{subequations} \label{eq:burgers}
\begin{gather}
    \frac{\partial u(x,t)}{\partial t} + u(x,t) \frac{\partial u(x,t)}{\partial x} = \nu  \frac{\partial^{2} u(x,t)}{\partial x^{2}} ,\label{eq:burgers_gen} \\
    u(x,0) = - \sin(\pi x),  \label{eq:burgers_ic} \\
    u(-1,t) = u(1,t) = 0, \label{eq:burgers_bc}
\end{gather}  
\end{subequations}
where the non-linear differential operator from \eqref{eq:pde} is $N[u(x,t)] = u(x,t) \frac{\partial u(x,t)}{\partial x} - \nu  \frac{\partial^{2} u(x,t)}{\partial x^{2}}$. The viscosity $\nu \in \{0.01, 0.05\}$ measures a fluid's resistance to flow. The initial condition \eqref{eq:burgers_ic} describes a sinusoidal shock at $t=0$ and the Dirichlet boundary condition \eqref{eq:burgers_bc} enforces the resting velocity at the bounds of $\Omega$. Figure \ref{fig:burgers_numerical} shows the solution to the Burgers' equation with a viscosity of $\nu=0.05$, which will be used for the main comparison of the physics-informed approach with the non-informed alternative.

\subsubsection*{Data}
\label{sec:sim_details_update}

\begin{figure}[ht]
    \centering
    \includegraphics[width=1\linewidth]{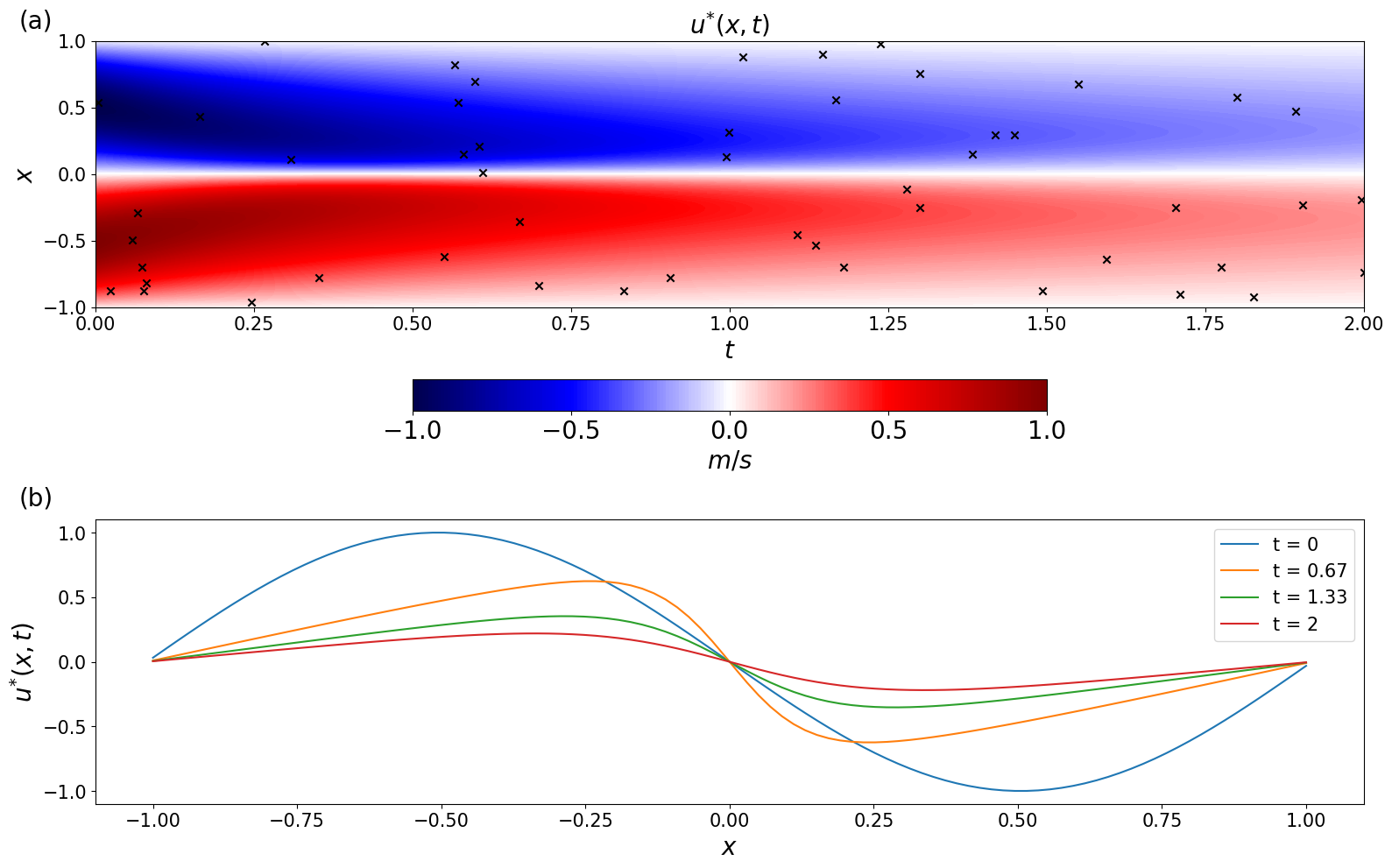}
    \caption{Solution $u^{*}(x,t)$ to the Burgers' equation in \eqref{eq:burgers}, with $\nu=0.05$. Panel (a) shows $u^{*}(x,t)$ throughout the domain \eqref{eq:burgers_domain} and the $n$ spatio-temporal locations chosen for training the model. Panel (b) shows $u^{*}(x,t)$ at four cross sections of $t \in \{0,0.67,1.33,2\}$. As time increases, the initial sinusoidal shock dissipates and the fluid reverts back to its resting state.}
\label{fig:burgers_numerical}
\end{figure}

We obtained solutions to the Burgers' equation using the \textit{PhiFlow} open-source simulation software (publicly available at  \url{github.com/tumpbs/PhiFlow}). We discretized the domain \eqref{eq:burgers_domain} with a $N_{s} \times N_{t}$ grid, where $N_{s}=100$ are points equally spaced in the spatial domain $[-1,1]$ and $N_{t}=1{,}000$ are points equally spaced in the temporal domain $[0,2]$, creating a grid of $N=N_{s} \times N_{t}$ points (as shown in Figure S1 of the supplementary material), and solved for $u^{*}(x_{i},t_{i})$, $i \in \{1, \ldots, N\}$, where $(x_i,t_i)$ are elements of the grid. In order to have more realistic data, we add Gaussian noise with variance $\sigma^{2}_{D}=0.1^{2}$,
\begin{gather}
    u(x_{i},t_{i}) = u^{*}(x_{i},t_{i}) + \epsilon_{i},~ 
    \epsilon_{i} \overset{\text{i.i.d.}}{\sim} N(0, \sigma_{D}^{2}), \label{eq:add_noise}
\end{gather}
so that $u(x_{i},t_{i})$ are $N$ noisy observations from the Burgers' equation model. We subsample $n=50$ data points uniformly in $\{u(x_{1},t_{1}),\ldots,u(x_{N},t_{N})\}$, shown in black in Figure \ref{fig:burgers_numerical}, and we assume we observe $\bm{u} = \{u(x_{1},t_{1}), \ldots, u(x_{n},t_{n})\}$ to predict the spatio-temporal process $u(x,t)$ across the entire domain of the Burgers' equation using our deep BNN model.

\subsubsection*{Physics-informed priors}
\label{sec:sim_details_priors}

As outlined in Section \ref{sec:methods}, the first step to implement our approach is to calibrate the prior distribution to a BNN using the physical model (\ref{eq:burgers}). We consider a BNN comprised of $L=4$ hidden layers, $k=10$ nodes in each layer, and a hyperbolic tangent  activation function $g(\cdot) = \tanh(\cdot)$. We employ the same $N_{s} \times N_{t}$ discretization grid of the domain \eqref{eq:burgers_domain} to assign prior means for $p(\bm{\theta})$. Denote with $N_{\text{PDE}} = N_{s} \times N_{t}$ the grid for the domain of \eqref{eq:burgers_gen}, $N_{\text{IC}}=N_{s}$ the grid for the domain of \eqref{eq:burgers_ic}, and $N_{\text{BC}}=2N_{t}$ the grid for the domain of \eqref{eq:burgers_bc}, for a total of  $N_{\text{PDE}}+N_{\text{IC}}+N_{\text{BC}}$ calibration points. We find $\hat{\bm{\mu}}_{\text{physics}}$ by minimizing
\begin{align*}
    \text{MSE}(\bm{\mu}) &\approx \frac{1}{N_{\text{PDE}}} \sum_{i=1}^{N_{\text{PDE}}} \left|\hat{f}_{\bm{\mu}} \left( x_{i},t_{i}\right) \right|^{2} + \frac{1}{N_{\text{IC}}} \sum_{i=1}^{N_{\text{IC}}} \left|\sin(\pi x_{i}) + \hat{u}_{\bm{\mu}}\left(x_{i},0\right) \right|^{2} \\
    &\qquad+ \frac{1}{N_{\text{BC}}} \sum_{i=1}^{N_{\text{BC}}} \left|\hat{u}_{\bm{\mu}}\left(-1,t_{i}\right) \right|^{2} + \frac{1}{N_{\text{BC}}} \sum_{i=1}^{N_{\text{BC}}} \left|\hat{u}_{\bm{\mu}}\left(1,t_{i}\right) \right|^{2}, \\ 
    &\hat{f}_{\bm{\mu}}(x_{i},t_{i}) := \frac{\partial \hat{u}_{\bm{\mu}}(x_{i},t_{i})}{\partial t} + \hat{u}_{\bm{\mu}}(x_{i},t_{i}) \frac{\partial \hat{u}_{\bm{\mu}}(x_{i},t_{i})}{\partial x} - \nu  \frac{\partial^{2} \hat{u}_{\bm{\mu}}(x_{i},t_{i})}{\partial x^{2}}.
\end{align*}
For simplicity, we set the prior variance as $\bm{\Sigma}_{l} = \sigma_{\text{physics}}^{2} \bm{I}$, with $\sigma_{\text{physics}}^{2}=0.0025^{2}$, for every $l$. The sensitivity of our approach's performance with respect to $n$, $\sigma^{2}_{\text{physics}}$, $\sigma^{2}_{D}$, the initial and boundary conditions, and the viscosity $\nu$ is addressed in Section \ref{sec:sim_sensitivity}.
 
\subsubsection*{Bayesian inference}
\label{sec:sim_details_inference}
Using the $n=50$ observations in the vector $\bm{u}$, we performed the BBB optimization (Algorithm \ref{alg:bbb}) with $N_{\text{BBB}}=10{,}000$ iterations to approximate the true posterior. The choice to rely on a small number of observation $n$ is intentional, as our approach is most effective for applications where data may be scarce. As such, using too many observations may defeat the purpose of relying on physics-informed priors. We performed $N_{\text{sim}}=100$ different simulations, each corresponding to a different realization of \eqref{eq:add_noise}. 

\subsubsection*{Performance metrics}
\label{sec:sim_details_metrics}

We measure each simulation's performance by reporting the mean squared error,
defined as the sum of the squared bias between the posterior mean and ground truth and the average posterior variance,
\begin{align}
    \text{MSE} = \frac{1}{N_{s} \times N_{t}}\sum_{i=1}^{N_{s}} \sum_{j=1}^{N_{t}} \left\{ \underbrace{\left[  \mathbb{E}\left(\mu_{u}(x_{i},t_{j}) \mid \bm{u}\right) - u^{*}(x_{i},t_{j}) \right]^{2}}_{\text{Bias}^{2}} + \underbrace{\mathbb{E}\left[\sigma_{u}^{2}(x_{i},t_{j}) \mid \bm{u}\right]}_{\text{Variance}} \right\}, \label{eq:mse_metric}
\end{align}
where $\mathbb{E}\left(\mu_{u}(x_{i},t_{j}) \mid \bm{u}\right)$ and $\mathbb{E}\left(\sigma_{u}^{2}(x_{i},t_{j}) \mid \bm{u}\right)$ are obtained as in \eqref{eq:post_mean}. 

\subsection{Comparing physics-informed and non-informed priors}
\label{sec:sim_results}

\begin{figure}[!ht]
    \centering
    \includegraphics[width=1\linewidth]{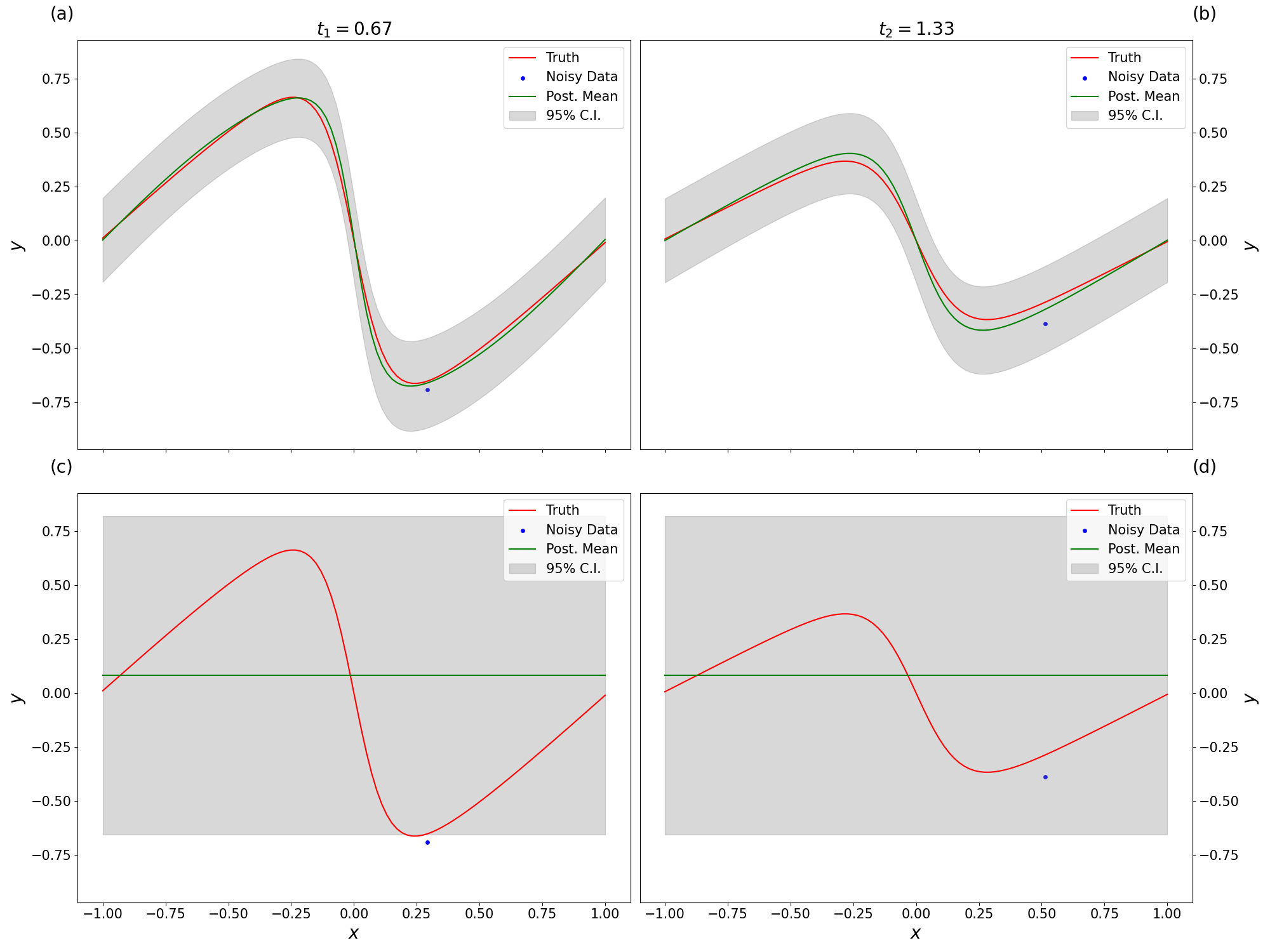}
    \caption{Comparison between physics-informed (a-b) and non physics-informed (c-d) predictions using the Burgers' equation at $t_{1}=0.67$ and $t_{2}=1.33$ (two of the same times from Figure \ref{fig:burgers_numerical}). Panels (a-b) show the simulation results using priors informed by the Burgers' equation \eqref{eq:burgers}. The domain-wide median squared bias and variance across the $N_{\text{sim}}=100$ simulations are $0.15 \times 10^{-2}$ and $1.08  \times 10^{-2}$, respectively. Panels (c-d) show results using a non-informed prior, now with $\bm{\mu}=\bm{0}$. In this case, the median squared bias and variance are $15.69  \times 10^{-2}$ and $16.73 \times 10^{-2}$, respectively.}
\label{fig:sim_summary}
\end{figure}

Table \ref{table:seeds_test} shows summary statistics across the $N_{\text{sim}}$ simulations, reporting the median metrics from the end of Section \ref{sec:sim_details_metrics} and their inter-quantile ranges. We just report the median computational demand, since it is practically the same across the $N_{\text{sim}}$ simulations. In the physics-informed case, we let the prior means be informed by the Burgers' equation, $\mathbb{E}(\bm{\theta}) = \hat{\bm{\mu}}_{\text{physics}}$ from \eqref{eq:mse} and $\bm{\Sigma}=\sigma^{2}_{\text{physics}}\bm{I}$, while the non-informed priors assume $\mathbb{E}(\bm{\theta}) = \bm{0}$ and $\bm{\Sigma}=\bm{I}$. A visual illustration of the difference in performance is shown in Figure \ref{fig:sim_summary}, where we show forecasts at two time points for one of the simulations. Panels (a-b) show simulation results using the physics-informed priors, while panels (c-d) show the same results using non-informative priors. 

\begin{table}[ht]
\centering
\setlength{\extrarowheight}{4pt}
\begin{tabular}{|c c| c|}
\hline
\multicolumn{1}{|c}{} & \multicolumn{1}{c}{Physics-Informed Priors} & \multicolumn{1}{c|}{Non-Informed Priors}  \\
Bias$^{2}$ $\left(10^{-2}\right)$ & $0.15~(0.12,0.19)$ & $15.69~(15.54,15.94)$  \\ 
Variance $\left(10^{-2}\right)$ & $1.08~(0.91,1.26)$ & $16.73~(14.11,18.47)$    \\
MSE $\left(10^{-2}\right)$ & $1.25~(1.06,1.42)$ & $32.36~(29.81,34.34)$  \\ \hline
Runtime (mm:ss) & $02:54$ & $02:58$  \\
 \hline
\end{tabular} 
\caption{Physics-informed vs. non-informed priors simulation using $n=50$ observations of the Burgers' equation \eqref{eq:burgers} with viscosity $\nu=0.05$. Displayed are the median and inter-quartile ranges across $N_{\text{sim}}=100$ different realizations for the bias, variance, and MSE as defined in equation \eqref{eq:mse_metric}. The computational demand does not materially change across the simulations, so we just report the median value.} \label{table:seeds_test}
\end{table}

The physics-informed model is able to retrieve a good approximation for the true solution (shown in red in Figure \ref{fig:sim_summary}) with a median squared bias of $0.15 \times 10^{-2}$, compared to an increase of almost two orders of magnitude across the non-informed simulations ($15.69 \times 10^{-2}$) for the non-informed case. This model-based approach provides \textit{informed} prior beliefs, while the scattered data points steer the posterior in the direction of observed behavior. As Figure \ref{fig:sim_summary}(a-b) shows, this knowledge propagates to the posterior yielding an improved forecast. Figure \ref{fig:sim_summary}(c-d) instead illustrates how the non-informed forecast is essentially a flat line.

Additionally, even though only $n=50$ observations were employed in the Bayesian update, the credibility intervals around the posterior mean (in gray in Figure \ref{fig:sim_summary}) are small relative to ground truth, as the median posterior variance across the domain is $1.08 \times 10^{-2}$, whereas in the non-informed case it rises by an order of magnitude ($16.73 \times 10^{-2}$). Notably, the median posterior variance in the physics-informed simulations resembles the amount of Gaussian noise we added, as $1.08 \times 10^{-2}$ is approximately $\sigma^{2}_{D}=0.1^{2}$. The posterior variance is a function of the confidence in the physical model, in this case $\sigma_{\text{physics}}^{2}=0.0025^{2}$, and the noise in the data, which we set at $\sigma^{2}_{D}=0.1^{2}$ and in the next section we perform a sensitivity analysis with respect to these quantities. 

\subsection{Sensitivity analysis}
\label{sec:sim_sensitivity}
We conduct a sensitivity analysis to understand our approach's performance as we change data availability or the viscosity in the Burgers' equation. For each scenario, we perform $N_{\text{sim}}=20$ different simulations and report the median MSE as defined in \eqref{eq:mse_metric} as well as the inter-quantile ranges. We report the median computational demand, since it does not change materially across the $N_{\text{sim}}$ simulations.  All of the simulations in the sensitivity study assume physics-informed priors. 

\subsubsection*{Sample size, spatial and temporal sampling, and noise}

We perform the same experiment with less and more observations ($n=2$, $250$, as opposed to $n=50$). We also study the effects of spatial and temporal subsampling. Finally, we test our approach for different amounts of Gaussian noise, $\sigma_{D}^{2}=0.5^{2}$ and $\sigma^{2}_{D}=0.001^{2}$, added to the solutions to the Burgers' equation. Beyond the sensitivity of our model performance, these simulations could replicate several scenarios of data availability such as the volume of collectable observations, the locations where sensors may be placed, or the reliability of those sensors. In Table \ref{table:sensitivity_results1}, we consider differing numbers of observations $n$ for the Bayesian update (in panel (a)), restrictions on the temporal domain in which observations are available (b), restrictions on the spatial domain in which observations are available (c), and magnitude of the Gaussian noise $\sigma_{D}^{2}$ (d).

As apparent from Table \ref{table:sensitivity_results1}(a), if $n=250$ observations are available, we observe a very low median bias but a larger median posterior variance compared to the scenario with $n=2$. When the number of observations is low, the Gaussian noise's influence on the posterior variance is minimal and the prior confidence in the PDE, $\sigma^{2}_{\text{physics}}=0.0025^{2}$, is the main determinant of the posterior variance. When instead $n=250$, the posterior variance mimics the magnitude of the Gaussian noise $\sigma_{D}^{2}$, as in the physics-informed results of Table \ref{table:seeds_test}. The simulations in Table \ref{table:sensitivity_results1}(b) assumes observations over a selected subset of the temporal domain, while the simulations in  Table \ref{table:sensitivity_results1}(c) assume that one can collect $n=50$ observations at just three and five locations across the spatial domain. The bias and variance in (b) are comparable to those of Table \ref{table:seeds_test}, which suggests that temporal sampling does not affect our approach's forecasting ability. Panel (c) instead shows a notable increase in bias when we restrict the spatial domain to just three locations compared to five, suggesting that collecting many observations at scattered times may be preferable to collecting many observations in the same locations over time (visual evidence is in Figures S2-S3 of the supplementary material). In Table \ref{table:sensitivity_results1}(d) we also see how increasing the Gaussian noise added to the solutions to the Burgers' equation by a factor of 5 still results in a good performance for the physics-informed prior, as it allows the posteriors to overcome the comparatively low signal-to-noise ratio in the data, yielding forecasts with low bias. At the same time, the amount of noise in the data propagates through the model, yielding a large posterior variance, as is apparent in Figures S4-S5 of the supplementary material. Overall, Table \ref{table:sensitivity_results1}(a-d) shows that, if the Bayesian priors are physics-informed and data are available at several spatial locations, our approach does not require a large number of observations to produce physically-consistent forecasts.

\begin{table}[ht]
\centering
\setlength{\extrarowheight}{2pt}
\scalebox{0.85}{
\begin{tabular}{|c c | c || c | c |}
\hline
\multicolumn{1}{|c}{} & \multicolumn{2}{c }{(a) \# of observations}  & \multicolumn{2}{c |}{(b) Temporal sampling} \\
\multicolumn{1}{|c}{} & $n=2$ & $n=250$ & $t \in[0,0.25]$ & $t \in[0.6,0.90]$ \\ \hline 
Bias$^{2}$ $\left(10^{-2}\right)$ & $3.05~(0.98,10.88)$ & $0.12~(0.11,0.13)$ & $0.22~(0.19,0.25)$ & $0.17~(0.14,0.20)$ \\ 
Variance $\left(10^{-2}\right)$ & $0.03~(0.02,0.10)$ & $1.16~(1.07,1.22)$ & $1.00~(0.87,1.27)$ & $1.02~(0.80,1.12)$   \\
MSE $\left(10^{-2}\right)$ & $3.18~(1.01,10.91)$ & $1.28~(1.19,1.33)$ & $1.22~(1.11,1.46)$ & $1.20~(1.02,1.30)$   \\ \hline
Runtime (mm:ss) & $00:53$ & $12:13$ & $04:16$ & $04:10$  \\ \hline \hline 
\multicolumn{1}{|c}{} & \multicolumn{2}{c }{(c) Spatial sampling}  & \multicolumn{2}{c |}{(d) Gaussian noise}  \\
\multicolumn{1}{|c}{} & $x \in \{0,0.5,1\}$ & $x \in \{-1,-0.5,0,0.5,1\}$ & $\sigma^{2}_{D}=0.5^{2}$ & $\sigma^{2}_{D}=0.001^{2}$  \\
\hline
 Bias$^{2}$ $\left(10^{-2}\right)$ & $3.28~(1.30,7.66)$ & $0.18~(0.13,0.23)$ & $1.06~(0.59,1.55)$ & $0.12~(0.11,0.13)$\\ 
 Variance $\left(10^{-2}\right)$ & $2.01~(1.24,10.49)$ & $0.99~(0.80,1.17)$ & $23.95~(21.48,28.87)$ & $0.11~(0.09,0.14)$   \\
 MSE $\left(10^{-2}\right)$ & $11.00~(4.58,16.81)$ & $1.18~(1.00,1.39)$ & $26.49~(22.47,30.60)$ & $0.23~(0.21,0.26)$   \\ \hline
 Runtime (mm:ss) & $03:24$ & $03:58$ & $04:10$ & $04:10$ \\
 \hline \hline
 \multicolumn{1}{|c}{}  & \multicolumn{2}{c }{(e) Prior variance} & \multicolumn{2}{c |}{(f) Initial and boundary conditions} \\
 \multicolumn{1}{|c}{}  & \multicolumn{2}{c }{} & \multicolumn{2}{c |}{and viscosity} \\
\multicolumn{1}{|c}{} &  &  & $u(x,0) = \cos (\pi x)$ &  \\
\multicolumn{1}{|c}{} & $\sigma^{2}_{\text{physics}}=0.01^{2}$ & $\sigma^{2}_{\text{physics}}=0.0001^{2}$ & $u(-1,t) = u(1,t)$ & $\nu=0.01$ \\ \hline
 Bias$^{2}$ $\left(10^{-2}\right)$ & $0.15~(0.13,0.18)$ & $0.14~(0.13,0.17)$ & $ 0.16~(0.14,0.19)$ & $0.19~(0.18,0.26)$ \\ 
 Variance $\left(10^{-2}\right)$ & $1.34~(1.24,1.51)$ & $1.06~(0.98,1.23)$  & $1.10~(0.93,1.22)$ & $1.15~(1.05,1.35)$ \\
 MSE $\left(10^{-2}\right)$ & $1.47~(1.43,1.71)$ & $1.19~(1.13,1.40)$  & $1.29~(1.10,1.50)$ & $1.38~(1.24,1.65)$ \\ \hline
 Runtime (mm:ss) & $04:11$ & $04:09$ & $03:54$ & $03:50$ \\
 \hline
\end{tabular}}
\caption{Sensitivity analyses of the physics-informed prior with respect to (a) the number of available observations, (b) the temporal sampling, (c) the spatial sampling, (d) the amount of Gaussian noise added, (e) the confidence in the physical model, and (f) the initial and boundary conditions and the viscosity in the PDE. Displayed are the median and inter-quartile ranges across $N_{\text{sim}}=20$ different realizations for the bias, variance, and MSE as defined in equation \eqref{eq:burgers}. The computational demand does not materially change across the simulations, so we just report the median value.} \label{table:sensitivity_results1}
\end{table}

\subsubsection*{Prior confidence in the PDE, initial and boundary conditions, and viscosity}   
    
Table \ref{table:sensitivity_results1}(e-f) shows results for different levels of prior confidence in the PDE as well as different initial and boundary conditions and viscosity. We test our approach under different levels of confidence in the PDE, $\sigma^{2}_{\text{physics}}=0.01^{2}$ and $\sigma^{2}_{\text{physics}}=0.0001^{2}$, a greater and smaller prior variance than in the simulations discussed in Section \ref{sec:sim_results} in Table \ref{table:sensitivity_results1}(e). We also consider a less viscous fluid, $\nu=0.01$, and a specification of the Burgers' equation with initial condition $u(x,0)=\cos(\pi x)$ and a periodic boundary condition, in Table \ref{table:sensitivity_results1}(f). 

Higher levels of confidence in the physical model, i.e., lower prior variance, translate to more accurate forecasts and lower levels of uncertainty, and vice-versa, as we show in Table \ref{table:sensitivity_results1}(e). When the prior variance is very low, our approach yields the smallest posterior bias, with an inter-quartile range of (0.13, 0.17), visually illustrated in Figure S6 of the supplementary material. On the other hand, lesser confidence in the physical model, in this case by a factor of 16, implies that a greater level of uncertainty propagates through the model, yielding greater posterior bias and variance. In Table \ref{table:sensitivity_results1}(f) show results from simulations whose priors are calibrated with respect to different initial and boundary conditions or viscosity, yielding a similar performance to the main simulation results we report in Table \ref{table:seeds_test}, as also shown in Figures S7-S8 of the supplementary material. 

\section{Application To Boundary Layer Flow}
\label{sec:ns_app}

We apply our approach to model and forecast boundary layer velocity from a high-resolution experimental data set from a water tunnel \citep{data1, data2}. Boundary layer velocity presents highly non-linear patterns, in time and space, described by the three-dimensional NS equations. Forecasting boundary layers presents considerable challenges, as the velocity field interacts with surfaces, and is one of the most active areas of research in fluid mechanics \citep{ml_reynolds, ml_reynolds2}, with applications in drag optimization along a moving object in fields such as aerospace, civil, or mechanical engineering. The section proceeds as follows. Section \ref{sec:ns_data} presents the data. Section \ref{sec:ns_priors} describes the physical model used to calibrate the physics-informed priors and the associated calibration of the prior mean. Section \ref{sec:ns_results} presents the results.

\subsection{Data}
\label{sec:ns_data}

The dataset used in this work is part of an experimental campaign to study boundary layer velocity focused near the bottom wall of a water tunnel. Data was collected in a cross-sectional, wall-normal plane parallel to the direction of the flow. The region of interest, in the middle of the 8-meter base, has a length of about 20 $cm$ along the wall, a height of about 5 $cm$, and a temporal resolution of 0.001 seconds over a total period of 5.004 seconds \citep{data1, data2}. The velocity data was obtained via Particle Image Velocimetry (PIV), the non-invasive optical measurement technique of choice when fluid mechanics experiments are performed with an unobstructed view of the flow \citep{piv}. PIV takes multiple laser-fluoresced images of a two-dimensional area of interest, spaced a known time interval apart, and calculates the fluid velocity on a Cartesian grid via statistical correlation of the seed particle locations between image pairs \citep{piv1, piv2, piv3, piv4}. 

The data is comprised of the \textit{instantaneous} velocity in the stream-wise and wall-normal directions (\textit{not} in the third, spanwise dimension), $u(\bm{s}_{i},t)$ and $v(\bm{s}_{i},t)$, respectively, in a $N=N_{x} \times N_{y}$ rectangular grid, $N_{x}=503$ and $N_{y}=124$, where $\bm{s}_{i}=(x_{i},y_{i})$ denotes the spatial coordinates $\{\bm{s}_{1},\ldots,\bm{s}_{N}\}$ and $t=1,\ldots,T=5{,}004$. We decompose the instantaneous velocity at spatial location $\bm{s}_{i}$ and time $t$ into its expectation (time trend) and fluctuation and average it over the $T$ time steps. In the case of the stream-wise velocity, we have 
\begin{gather*}
    \frac{1}{T} \sum_{t=1}^{T} u(\bm{s}_{i},t) = \frac{1}{T} \sum_{t=1}^{T}  \left\{ \underbrace{\overline{u}(\bm{s}_{i})}_{\text{time trend}} + \underbrace{u'(\bm{s}_{i},t)}_{\text{fluctuation}} \right\} = \overline{u}(\bm{s}_{i}),
\end{gather*}
since the average of the fluctuations over time is zero by definition. The manipulation for the instantaneous wall-normal velocity $v(\bm{s},t)$ is similar. Since the data collection region is sufficiently downstream in the water tunnel, the flow can be considered fully developed --- i.e., gradients in the mean velocities are approximately zero. We also assume that there are no gradients in the mean velocities in the spanwise direction. With these assumptions, mass conservation, which requires a divergence-free mean velocity for an incompressible flow, implies $\partial \overline{v} / \partial y \approx 0$. At the lower wall, $\overline{v}(0,y)=0~m/s$, so that $\overline{v}(\bm{s}_{i}) = 0$, $i \in \{1,\ldots,N\}$. Thus we expect that the only non-zero mean velocity component $\overline{u}$ is approximately only a function of the wall-normal coordinate $y$. Note, however, that for generality, we retain an explicit dependence of $\overline{u}$ on both $x$ and $y$ (i.e., $\overline{u}(x,y) = \overline{u}(\bm{s}_{i}))$. Diagnostics are available in Figure S9 of the supplementary material. 

\begin{figure}[!ht]
    \centering
    \includegraphics[width=\linewidth]{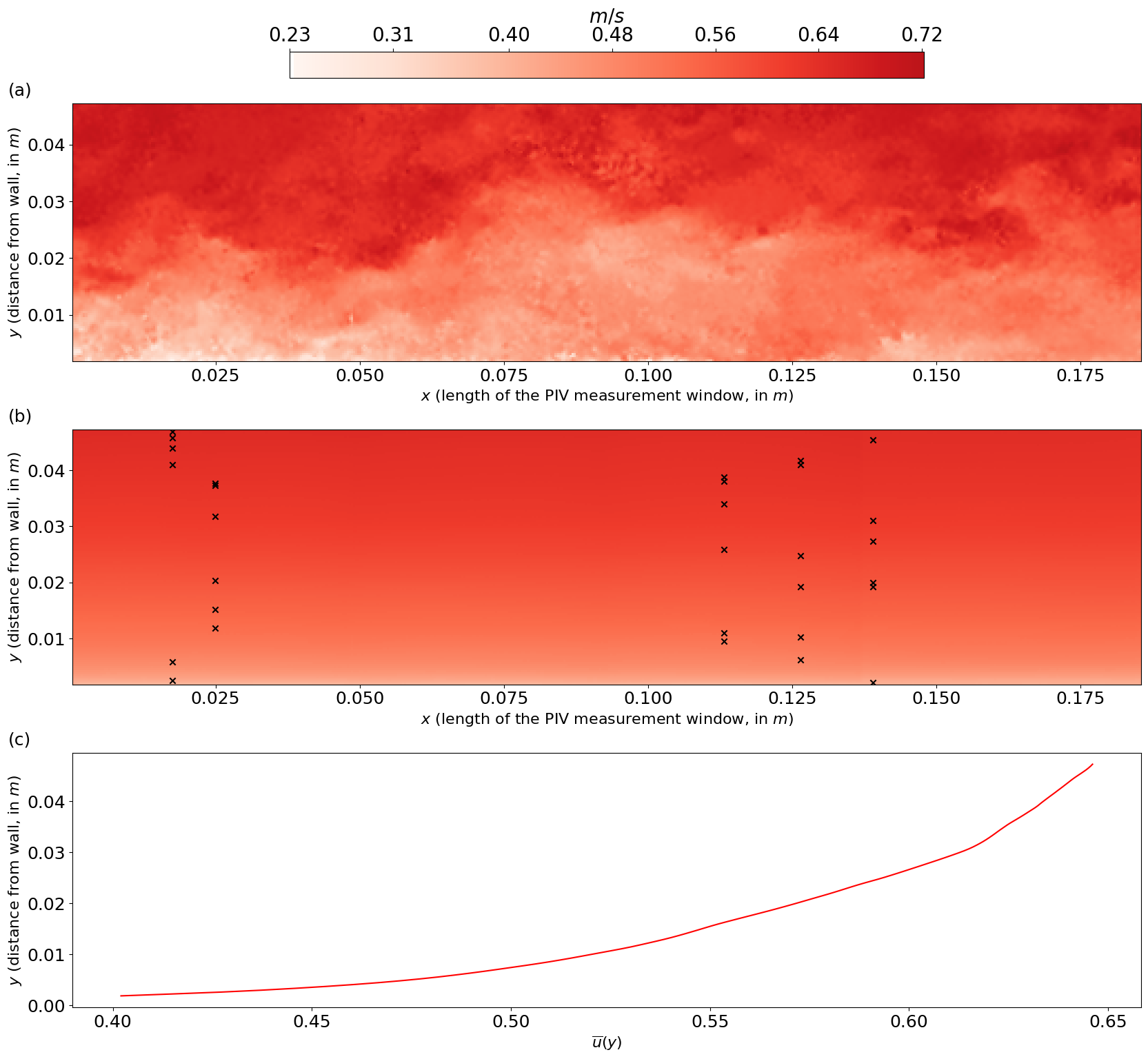}
    \caption{Boundary layer data collected in a rectangular window inside an experimental water tunnel. Panel (a) shows a single snapshot of the instantaneous stream-wise velocity (going from left to right), namely $t=,1{,}000$, $\bm{s}_{i} \in \{\bm{s}_{1},\ldots,\bm{s}_{N}\}$. Shown in panel (b) is the stream-wise velocity $\overline{u}(\bm{s}_{i})$, after averaging across the $T=5{,}004$ time steps. The pattern is stratified from the wall ($x \approx 0$) to the top of the domain, where $\overline{u}$ approaches the free-stream, i.e., undisturbed by the wall, velocity which the experiment set at 0.67 $m/s$. The points are a subsample of size $n=30$ used to predict the entire domain. Panel (c) displays the average vertical profile of $\overline{u}$, i.e., after averaging out $x$, showing the behavior of the stream-wise velocity in a fully-developed turbulent boundary layer.}
    \label{fig:data_ubar}
\end{figure}

In summary, we consider $\overline{u}(\bm{s}_{i})$, the average velocity in the stream-wise direction across the $5{,}004$ time steps. We subsample observations from five vertical cross sections and only keep 5\% of them from each section, as shown in black in Figure \ref{fig:data_ubar}(b), so the data used to predict the entire spatial domain are now $\bm{u} := \bm{\overline{u}}=\left\{\overline{u}(\bm{s}_{1}), \ldots, \overline{u}(\bm{s}_{n}) \right\}$, $n=30$, which amounts to less than 0.05\% of the $N$ total observations.

Figure \ref{fig:data_ubar}(a) shows one of the $T=5{,}004$ snapshots of the instantaneous stream-wise velocity going from left to right, in this case $u(\bm{s},1{,}000)$. The turbulent nature of this boundary layer is clearly shown in the large variance of the gradients in both the stream-wise and wall-normal directions. Figure \ref{fig:data_ubar}(b) shows the average stream-wise velocity $\overline{u}(\bm{s})$, where the spatial coordinates $x$ and $y$, both measured in meters, represent the base and height of the PIV measurement window, respectively. Panel (c) shows the average (in both time and $x$) vertical profile of $\overline{u}$ which varies from $0.4~m/s$ near the wall to 0.65 $m/s$ at the top of the observable domain. As we move away from the surface, the flow's velocity approaches its free-stream profile, i.e., undisturbed by the wall, which the experiment set at $0.67~m/s$. The lowest recorded velocity being $0.4~m/s$ (close to but \textit{not at} the wall) is consistent with boundary layer theory, as the gradient of $\overline{u}$ with respect to the wall-normal direction is largest near the wall, and we would expect measurements of $\overline{u}$ at $y=0$, if recordable, to be $0~m/s$ \citep{kundu}. Since the flow moves from left to right, time-averaging smooths out most of the vertical perturbations and the wall-normal component becomes negligible (see Figures \ref{fig:data_ubar}(a) and (b)). 

\subsection{Governing PDE}
\label{sec:ns_priors}

The (three-dimensional, instantaneous) velocity in turbulent boundary layers is governed by the NS equations, which encode momentum and mass conservation. The lack of observations in the third dimension compels us to consider the Reynolds-averaged (i.e., time-averaged) Navier-Stokes (RANS) equations, a static version of NS that governs the behavior of the stream-wise velocity in space. Under appropriate assumptions (verified for this data in the supplementary material), the governing model simplifies to a one-dimensional PDE (the full derivation is also in the supplementary material):
\begin{subequations} \label{eq:rans}
\begin{align}
    \frac{u_{\tau}^{2}}{\delta} + \frac{\delta u_{\tau}}{Re} \frac{\partial^{2} \overline{u}(x,y)}{\partial y^{2}} - \frac{\partial}{\partial y} \overline{u'v'} = 0, \label{eq:rans_gen} \\ 
    -\overline{u'v'} = \underbrace{(\kappa y) \frac{\overline{u}(x,y)}{u_{\tau}} \frac{\partial \overline{u}(x,y)}{\partial y}}_{\substack{\text{Reynolds stress} \\ \text{approximation}}}, \label{eq:closure}
\end{align} 
\end{subequations}
where \ref{eq:closure} introduces a standard model for the Reynolds stress term $\overline{u'v'}$. In this case the derivative of $\overline{u}$ with respect to time is zero and the non-linear differential operator as defined in the PDE \eqref{eq:pde_gen} is
\begin{gather*}
    N\left[\overline{u}(x,y)\right] = \frac{u_{\tau}^{2}}{\delta} + \frac{\delta \overline{u}_{\tau}}{\text{Re}} \frac{\partial^{2} \overline{u}(x,y)}{\partial y^{2}} + \frac{\partial}{\partial y} \left( (\kappa y) \frac{\overline{u}(x,y)}{u_{\tau}} \frac{\partial \overline{u}(x,y)}{\partial y} \right),
\end{gather*}
where from the experimental data, $\text{Re}=2{,}700$ is the Reynolds number, $\delta=0.1~m$ is the boundary layer thickness, $u_{\tau}=0.027~m/s$ is the friction velocity, and $\kappa=0.01107$. The boundary conditions are $\overline{u}(x,0) = 0~ m/s$ and $\overline{u}(x,\delta)=U_{\infty}$, where $U_{\infty}=0.67~m/s$ is the free-stream velocity.

\subsubsection*{BNN assumptions and prior calibration}

We employ a deep BNN comprised of $L=2$ hidden layers, $k=10$ nodes in each layer, a hyperbolic tangent  activation function $g(\cdot) = \tanh(\cdot)$, and the likelihood is now expressed as $\overline{u}(x,y) \sim \mathcal{N}(\mu_{\overline{u}}(x,y), \sigma^{2}_{\overline{u}}(x,y))$. We can afford a shallower BNN here than the one in Section \ref{sec:sim_study} because the process is static in time and the spatial relationships are one-dimensional, as shown in Figure \ref{fig:data_ubar}.

Since the boundary conditions on \eqref{eq:rans} are $\overline{u}(x,0)=0$ and $\overline{u}(x,\delta)=0.67$ and the data only covers the wall-normal direction up to $y=0.0473~m$, we discretize the interval $[0,\delta]$ with $N_{y}$ points and use a $N_{\text{PDE}} = N_{x} \times N_{y}$ grid ($N_{x}$ and $N_{y}$ are defined as in Section \ref{sec:ns_data}) to obtain $\hat{\bm{\mu}}_{\text{physics}}$ by minimizing the MSE with respect to \eqref{eq:rans}:

\begin{align}
    \text{MSE}(\bm{\mu}) &\approx \frac{1}{N_{\text{PDE}}} \sum_{i=1}^{N_{\text{PDE}}} \left|\hat{f}_{\bm{\mu}} \left( x_{i},y_{i}\right) \right|^{2} + \frac{1}{N_{x}} \sum_{i=1}^{N_{x}} \left\{\left|\hat{\overline{u}}_{\bm{\mu}}\left(x_{i},0\right) \right|^{2} + \left|\hat{\overline{u}}_{\bm{\mu}}\left(x_{i},\delta \right)-U_{\infty}\right|^{2} \right\}, \label{eq:ns_mse} \\ 
    &\hat{f}_{\bm{\mu}}(x_{i},y_{i}) := \frac{u_{\tau}^{2}}{\delta} + \frac{\delta u_{\tau}}{Re} \frac{\partial^{2} \hat{\overline{u}}_{\bm{\mu}}(x_{i},y_{i})}{\partial y^{2}} + \frac{\partial}{\partial y} \left\{(\kappa y_{i}) \frac{\hat{\overline{u}}_{\bm{\mu}}(x_{i},y_{i})}{u_{\tau}} \frac{\partial \hat{\overline{u}}_{\bm{\mu}}(x_{i},y_{i})}{\partial y} \right\}.\nonumber
\end{align}
Figure \ref{fig:rans_main_fig}(c) in Section \ref{sec:ns_results} includes the results of the prior calibration (in blue), showing a clear departure from the observations (in red). Even with \eqref{eq:ns_mse} on the order of magnitude of $10^{-5}$, the inferred approximation $\hat{\overline{u}}_{\hat{\bm{\mu}}_{\text{physics}}}$ shows a clear discrepancy with $\overline{u}$, due to the Reynolds stress approximation in \eqref{eq:closure}. Our prior confidence in $\hat{\bm{\mu}}_{\text{physics}}$ is therefore smaller than in the simulation study in Section \ref{sec:sim_study} and we set $C=0.03$, where $C$ is as defined in Section \ref{sec:pinn_background}. By comparison, the average entry of $\bm{\Sigma}$, from \eqref{eq:priors_gen}, is over 4 times larger than it is in the simulation study, where each diagonal entry of $\bm{\Sigma}$ is equal to $\sigma^{2}_{\text{physics}}=0.0025^{2}$ . In Section \ref{sec:ns_results} we also consider $C=0.01/100$ to discuss the role of an excessively high prior confidence in \eqref{eq:rans}. 

Using the $n=30$ observations in the vector $\Bar{\bm{u}}$, we perform the Bayesian update with variational inference with $N_{BBB}=10{,}000$ iterations of Algorithm \ref{alg:bbb} to approximate the true posterior. This particular choice of observations is meant to replicate an instrument that can record two-dimensional velocity at scattered locations along a vertical line with some amount of measurement error. 

\subsection{Results}
\label{sec:ns_results}

\begin{figure}[!ht]
    \centering
    \includegraphics[width=\linewidth]{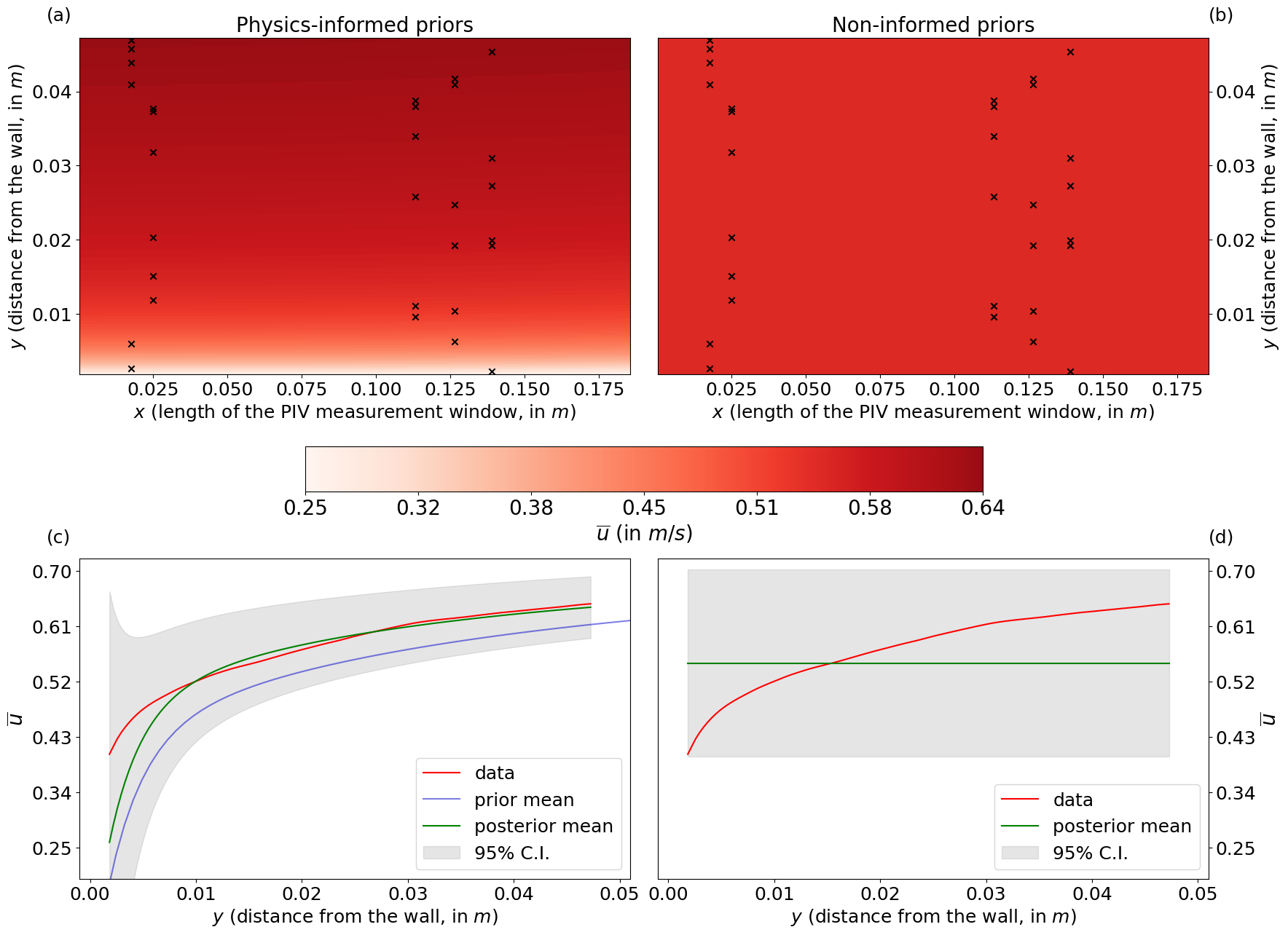}
    \caption{Comparing boundary layer predictions with physics-informed priors, panels (a) and (c), and non-informed priors, panels (b) and (d). The spatial forecast for each is shown in panels (a-b) along with the $n=30$ observations. Panels (c-d) show the average vertical profile of the predicted and observed velocity, along with the 95\% credibility intervals. In panel (c), we also show the velocity forecast induced by the prior means. We do not show its counterpart in panel (d) since it is a flat line at $0~m/s$.}
    \label{fig:rans_main_fig}
\end{figure}

We applied our approach with a physics-informed prior, $\mathbb{E}(\bm{\theta}) = \hat{\bm{\mu}}_{\text{physics}}$ and $\bm{\Sigma}$ determined by $C\in\{0.03,~0.01/100\}$, as well as a non-informed prior, $\mathbb{E}(\bm{\theta}) = \bm{0}$ and $\bm{\Sigma}_{l}=(1/N_{l})\bm{I}$, where $N_{l}$ is the number of parameters in the $l$-th layer. Figures \ref{fig:rans_main_fig}(a-b) show the forecast throughout the spatial domain and Figures \ref{fig:rans_main_fig}(c-d) show the average vertical profiles of the stream-wise velocity and 95\% credibility intervals. A similar figure for the case with $C=0.01/100$ is available in the supplementary material. Table \ref{table:rans_main_table}(a) and \ref{table:rans_main_table}(c) present the MSE as introduced in \eqref{eq:mse_metric}.

Our results show a clear improvement in performance both in terms of bias and variance when the priors are informed by the PDE in \eqref{eq:rans}. The bias and variance in the physics-informed case are $0.77 \times 10^{-3}$ and $43.38 \times 10^{-3}$, respectively, as the spatial map in Figure \ref{fig:rans_main_fig}(a) shows the forecast velocity in a similar stratified pattern to the data in Figure \ref{fig:data_ubar}. The non-informed model forecasts a flat stream-wise velocity, approximately equal to the wall-normal average for all $y$, but is unable to capture the non-linearity in the profile, as is apparent in Figure \ref{fig:rans_main_fig}(c-d), with a bias and variance equal to $4.20 \times 10^{-3}$ and $77.66 \times 10^{-3}$, respectively. Most importantly, boundary layers such as the one outlined in Section \ref{sec:ns_data} always exhibit increasing gradients in the wall-normal direction \citep{kundu} and a forecast such as the one shown in Figure \ref{fig:rans_main_fig}(b) is not only poor but also \textit{unphysical}. On the other hand, the physics-informed forecast in Figure \ref{fig:rans_main_fig}(a) shows a physically-consistent boundary layer due to the spatial relationship established a priori.

Additional evidence for the improvement in performance in the physics-informed case is in a diagnostic analysis of the Reynolds stress $\overline{u'v'}$, which, physically, represents momentum transport due to covariances in the turbulent velocity fluctuations. From the instantaneous velocity fields $u(\bm{s},t)$ and $v(\bm{s},t)$, we computed $\overline{u'v'}$ for our data as
\begin{gather*}
    \overline{u'v'}(\bm{s}) = \sum_{t=1}^{T} u(\bm{s},t)v(\bm{s},t) - \left[\sum_{t=1}^{T} u(\bm{s},t)\right] \left[\sum_{t=1}^{T}v(\bm{s},t)\right],
\end{gather*}
and compared it with its estimated value (derived from \eqref{eq:closure}) by computing the MSE across the spatial domain, where, recall, the points are on an evenly-spaced grid. In the physics-informed case the MSE is $4.16 \times 10^{-7}$, one order of magnitude smaller than the MSE in the non-informed case, $1.53 \times 10^{-6}$, and visual evidence for this difference is in Figure S12 of the supplementary material.

The results with high prior confidence, $C=0.01/100$, show the role of excessive confidence in the prior. As can be seen in Table \ref{table:rans_main_table}(b), setting a high prior confidence in the PDE prevents the data from properly augmenting the physics-informed prior. In this case the bias increases by over 70\%, $1.31 \times 10^{-3}$, even though the variance is lower, $12.96 \times 10^{-3}$. The governing PDE \eqref{eq:rans} does \textit{not} faithfully describe observed behavior in the way that the Burgers' equation \eqref{eq:burgers} did in Section \ref{sec:sim_study}, since the Reynolds stress approximation in \eqref{eq:closure} is only approximately correct. As such, \eqref{eq:rans} constitutes prior knowledge on the behavior of $\overline{u}$, but only loosely so. The physics-informed model with very high prior confidence unbalances the posterior in the direction of the PDE and, while outperforming the non-informed model, falls short of producing a reliable forecast.  

\begin{table}[!ht]
\centering
\setlength{\extrarowheight}{4pt}
\scalebox{1}{
\begin{tabular}{|c c| c||c |}
\hline
\multicolumn{1}{|c}{} & \multicolumn{2}{c}{Physics-Informed Priors} & \multicolumn{1}{c|}{Non-Informed Priors} \\
& $C=0.03$ & $C=0.01 / 100$ & $\bm{\Sigma}_{l} = (1/N_{l})\bm{I}$  \\ 
& (a) & (b) & (c)  \\ \hline
Bias$^{2}$ $\left(10^{-3}\right)$ & $0.77$ & $1.31$ & $4.20$ \\ 
Variance $\left(10^{-3}\right)$ & $43.38$ & $12.96$ & $77.66$   \\
MSE $\left(10^{-3}\right)$ & $44.15$ & $14.27$ & $81.86$ \\ \hline
Runtime (mm:ss) & $05:54$ & $05:58$ & $06:10$ \\
 \hline
\end{tabular}}
\caption{Physics-informed application results of $n=30$ observations for different levels of prior confidence, as defined in Section \ref{sec:pinn_background}, $C\in\{0.03, 0.01/100\}$, in panels (a-b). Panel (c) reports the results for the non-physics informed application, i.e., with a vague prior. Displayed are the bias, variance, and MSE as defined in Section \ref{sec:sim_details_metrics} for the predicted stream-wise velocity $\overline{u}$. We also report the computational demand.} \label{table:rans_main_table}
\end{table}

These results have significant practical implications. Fluid dynamics data sets are notoriously difficult to obtain and experimental settings are often required. Since we consider the time-averaged stream-wise velocity, the vast majority of the variability in the data is in the $y$ direction, as seen in Figure \ref{fig:data_ubar}. In spite of this simplification and the $n=30$ observations being collected along 5 \textit{vertical} cross sections of the domain, the non-informed models were unable to identify the spatial relationships in the data. The physics-informed approach instead established the spatial dependence a priori and did not require observations evenly distributed in space to produce physically-consistent forecasts. 

\section{Conclusion}
\label{sec:conclusion}

This work introduced a novel model-based Bayesian approach to incorporate contextual knowledge in the form of a PDE in a BNN to forecast spatio-temporal processes. The proposed model produces improved forecasts relative to a BNN with vague priors, especially with a small amount of data, as we show in both a simulation study and an application to boundary layer velocity. We propose a different paradigm of PINNs by incorporating physical knowledge in the prior \textit{without} relying on observations, rather than expressing it as a constraint, so that true a priori knowledge is encoded. Given a level of confidence in the priors, expressed as a prior variance, the result is a physics-informed forecast. Incorporating in the prior the spatio-temporal relationships described by a PDE constitutes a modeling strategy to borrow strength from a governing model and data to produce physically-consistent forecasts.

In a simulated environment governed by the viscous Burgers' equation, our Bayesian framework shows strong predictive skills in terms of posterior mean squared error despite using a small number of observations. We also perform a sensitivity study and assessed how a larger number of observations, lower amounts of noise, and lower prior variance all lead to improved posteriors. 

We apply our approach to average boundary layer velocity as it interacts with a surface, governed by a simplified NS that only approximately describes observed behavior due to additional modeling assumptions.  The physics-informed prior again shows superior forecasting abilities to the vague prior alternative. Our results also show that an excessively high confidence in the PDE yields forecasts that resemble theoretical behavior too closely, as the posterior skews in the direction of the physics-informed prior means. Overall, the application results show that a governing model need not perfectly describe observed behavior in order for our approach to perform better than the vague prior alternative.

Our approach represents the first step towards a Bayesian-driven model-based methodology that yields physically-consistent predictions. In NNs, the parameters, even if physics-informed a priori, lack interpretability and measuring our prior confidence in them without employing observations is challenging. We introduce a metric inspired by the coefficient of variation so that each parameter's prior variance is relative to its mean, but it still does not easily translate to \textit{confidence}. As such, our results may be interpreted as conditional on a level of prior confidence, as our approach shows promising results in its ability to propagate uncertainty through the model. For a given level of noise in the data, the posterior variance in both the simulation study and the application follows the prior confidence, as is expected in any Bayesian model. 

An important issue to be addressed in future work concerns the physics with respect to which we calibrate the prior. The governing models employed in Sections \ref{sec:sim_study} and \ref{sec:ns_app} assume no uncertainty in the PDE itself, as our results may also be interpreted as conditional on the physical parameters. Large portions of the PINN literature focus on finding the best physical parameters for a given system \citep{raissi, b_pinn, e_pinn} and an extension of this work would impose a prior on them as well. Such a framework would provide, in principle, a more rigorous quantification of our level of prior confidence in the governing model while producing a posterior on the physical parameters. Finally, the flexibility of our approach, along with its modest computational demand, shows promise for a future application to the three-dimensional NS equations, which is more realistic than the average boundary layer velocity studied in Section \ref{sec:ns_app}. Researchers typically tackle these problems using computational fluid dynamics, which solve NS in every point of a fine mesh and whose computational demand is especially high for complex flows \citep{cfd_3d}. As the classical PINN approach has already shown promise towards easing the computational burden \citep{cfd_review}, extensions of our work may provide an alternative model-based strategy with significant implications in several areas of science and engineering. 

\section*{Code and Data Availability}

This code and data can be found in the following GitHub repository: \url{github.com/Env-an-Stat-group/25.Menicali.IEEETNNLS}. The application data that support the findings of this study are openly available upon request from \cite{data_website}.

\bibliography{refs}

\end{document}